\documentclass[preprint,showkeys,aps,prb]{revtex4}
\usepackage[dvips]{graphicx}

\begin{document}

\title{
2020 Ian Snook Prize Problem : Three Routes to the Information Dimensions for One-Dimensional
Stochastic Random Walks and Their Equivalent Two-Dimensional Baker Maps.
}

\author{
William Graham Hoover and Carol Griswold Hoover                    \\
Corresponding Author email : hooverwilliam@yahoo.com               \\
Ruby Valley Research Institute, 601 Highway Contract 60            \\
Ruby Valley Nevada 89833, USA ;     \\
}

\date{\today}


\vspace{0.1cm}

\begin{abstract}

The \$1000 Ian Snook Prize for 2020 will be awarded to the author(s) of the most interesting paper exploring
pairs of relatively simple, but fractal, models of nonequilibrium systems, dissipative time-reversible
Baker Maps and their equivalent stochastic random walks.  Two-dimensional deterministic, time-reversible,
chaotic, fractal, and dissipative Baker maps are equivalent to stochastic one-dimensional random walks.
Three distinct estimates for the information dimension, $\{ \ 0.7897,\ 0.741_5, \ 0.7337 \ \}$ have all been
put forward for one such model.  So far there is no cogent explanation for the differences among these estimates.
We describe the three routes to the information dimension, $D_I$: [ 1 ] iterated Cantor-like mappings, [ 2 ]
mesh-based analyses of single-point iterations, and [ 3 ] the Kaplan-Yorke Lyapunov dimension, thought by many
to be exact for these models. We encourage colleagues to address this Prize Problem by suggesting, testing, and
analyzing mechanisms underlying these differing results.  

\end{abstract}

\maketitle

\begin{figure}
\includegraphics[width=1.8 in,angle=-90.]{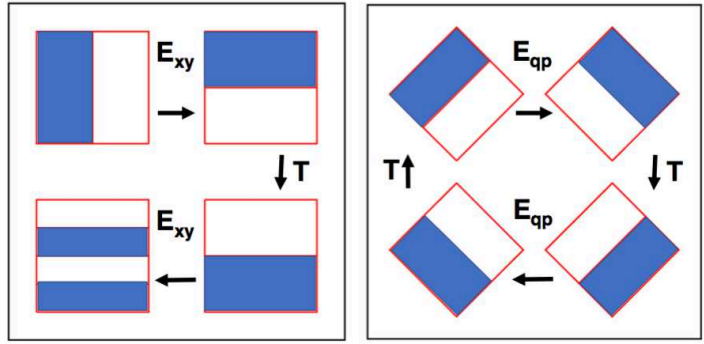}
\caption{
The actions of the Cartesian E$_{\rm xy}$ and Rotated E$_{\rm qp}$ equilibrium Baker Maps are shown here.
The $(q,p)$ map is ``time-reversible'' in the sense that the time-reversal mapping
T simply changes the sign of the surrogate momentum $p$, T$(\pm q,\pm p) = (\pm q,\mp p)$. Here $q$ is horizontal
and $p$ is vertical.  Both have extreme values of $\pm\sqrt{2}$.  Note that E$_{\rm xy}$ is irreversible.
}
\end{figure}

\begin{figure}
\includegraphics[width=1.8 in,angle=-90.]{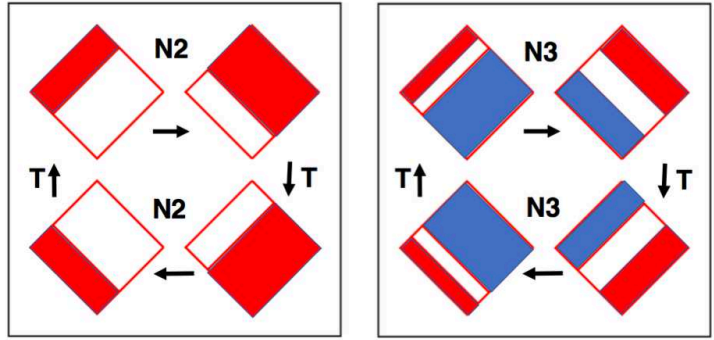}
\caption{
Actions of two Nonequilibrium generalizations of the Baker Map,
N2 and N3, are shown here. These $(q,p)$ maps are ``time-reversible'' in the sense that the time-reversal mapping
T simply changes the sign of the surrogate momentum $p$, T$(\pm q,\pm p) = (\pm q,\mp p)$. Here $q$ is horizontal
and $p$ is vertical.  Both have extreme values of $\pm\sqrt{2}$. The nonequilibrium attractors corresponding to
these two maps are shown in {\bf Figure 3}. These two $(q,p)$ maps have unstable fixed points at the tops and
bottoms of their diamond-shaped domains.
}
\end{figure}

\begin{figure}
\includegraphics[width=1.5 in,angle=-90.]{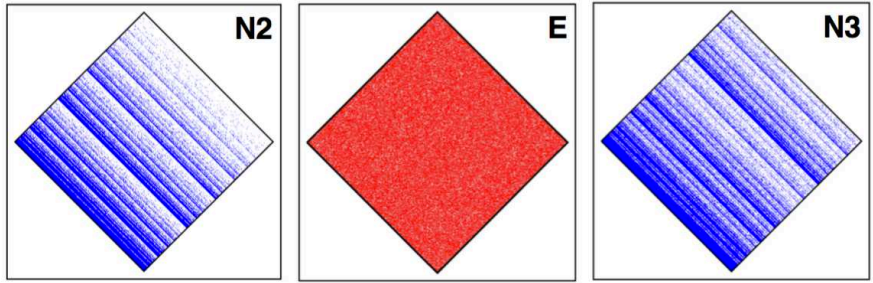}
\caption{
The iterated actions (200 000 points) of the Nonequilibrium generalizations N2 and N3 of the equilbrium Baker Map E
are shown here. The analytic forms of these maps in the $(q,p)$ coordinates of this figure are given in the text. The
two attractors at the right and left have similar information dimensions, but with precise values for N2 currently
uncertain. The 2020 Snook Prize problem is to shed light on this uncertainty. The equilibrium E$_{\rm qp}$ map
provides the homogeneous red covering of the phase space shown in the middle of the figure.
}
\end{figure}

\section{Introduction and Background}

In conjunction with the Institute of Bioorganic Chemistry of the Polish Academy of Sciences (at the Poznan
Supercomputing and Networking Center) we again offer our \$500 Snook Prize along with the Institute's \$500 Additional
Prize for the best paper addressing the upcoming 2020 year's Prize Problem.  Ian Snook's speciality was statistical
mechanics so that this memorial Prize is directed toward research efforts in or near to his chosen field.
Details of past problems and the award can be found online at the websites of Computational Methods in Science
and Technology, {\tt cmst.eu}, or at our own {\tt hooverwilliam.info}. 

\subsection{Lessons Learned from Hopf's Baker Map}

The upcoming year's 2020 Snook Prize problem involves a generalization of Eberhard Hopf's Baker map, discussed at a
well-attended seminal meeting on ``Chaos and Irreversibility'' at E\"otv\"os University-Budapest in 1977\cite{b1,b2}. Both 
equilibrium ( incompressible ) and two nonequilibrium ( with twofold area changes ) mappings are described in {\bf Figures
1 and 2}. Hopf's $(x,y)$ maps have been rotated $45^o$ to give $(q,p)$ forms satisfying ``time reversibility'': E$^{-1}$ =
T{\tt *}E{\tt *}T; N2$^{-1}$=T{\tt *}N2{\tt *}T, where ${\rm T}(q,p) = (q,-p)$.  To illustrate the rotation let us begin
with the conventional area-conserving Baker Map.  For convenience we define the mapping in a $2 \times 2$ area centered on
the origin:
\begin{verbatim}
if(x.le.0) xnew = 2*x + 1
if(x.le.0) ynew = (y + 1)/2
if(x.gt.0) xnew = 2*x - 1
if(x.gt.0) ynew = (y - 1)/2
[ Equilibrium xy Baker Map ]
\end{verbatim}
This {\tt (x,y)} form of the Equilibrium Baker Mapping can be converted to a more useful {\tt (q,p)} mapping using the
transformation 
$$
{\textstyle
q = \sqrt{\frac{1}{2}}(y+x) \ ; \ p = \sqrt{\frac{1}{2}}(y-x) \ \longleftrightarrow \ 
x = \sqrt{\frac{1}{2}}(q-p) \ ; \ y = \sqrt{\frac{1}{2}}(q+p) \ . 
}
$$

The changes ${\tt (x,y)} \to {\tt (q,p)}$ and ${\tt (xnew,ynew)} \to {\tt (qnew,pnew)}$ result in the $(q,p)$ form of
the Equilibrium Baker Mapping E, with the ``coordinate'' $q$ horizontal and the ``momentum'' $p$ vertical :
\begin{verbatim}
if(q-p.le.0) qnew = + (5/4)*q - (3/4)*p + 3*d
if(q-p.le.0) pnew = - (3/4)*q + (5/4)*p - 1*d
if(q-p.gt.0) qnew = + (5/4)*q - (3/4)*p - 3*d
if(q-p.gt.0) pnew = - (3/4)*q + (5/4)*p + 1*d
[ Equilibrium qp Baker Map ]
\end{verbatim}
We term these two ``equilibrium'' maps by analogy to Hamiltonian mechanics as both of them preserve area, $dxdy$ or
$dqdp$. The constant $d$, which determines the scale of the mapping, is equal to $\sqrt{1/8}$. The inverse mapping
E$^{-1} =$ T{\tt *}E{\tt *}T is similar :
\begin{verbatim}
if(q+p.le.0) qnew = + (5/4)*q + (3/4)*p + 3*d
if(q+p.le.0) pnew = + (3/4)*q + (5/4)*p + 1*d
if(q+p.gt.0) qnew = + (5/4)*q + (3/4)*p - 3*d
if(q+p.gt.0) pnew = + (3/4)*q + (5/4)*p - 1*d
[ Inverse Equilibrium Map ]
\end{verbatim}
The ``obvious" initial condition $(q,p)=(0,0)$ lies on the singular dividing lines of these maps and should be
avoided.  The choice {\tt (q,p) = (1,0)} is satisfactory.  Our single-precision simulations with this choice
settled onto periodic orbits of length 143,512 iterations.  Double-precision simulations have much longer 
periodic orbits, with trillions of iterations.  The same $(1,0)$ initial condition settled onto a periodic orbit
of length 3,412,524,575,046 iterations.  In all of our simulations we have used FORTRAN77 in programs
written as transcriptions of the pseudocoding shown here.

The ``nonequilibrium" Baker Mapping N2 incorporates area changes while remaining ergodic, covering the entire
$2\times 2$ phase space and with a comoving twofold change in area with each iteration.  This is N2 :
\begin{verbatim}
if(q-p.le.-sqrt(2/9)) qnew = + (11/ 6)*q - ( 7/ 6)*p + 14*d
if(q-p.le.-sqrt(2/9)) pnew = - ( 7/ 6)*q + (11/ 6)*p - 10*d
if(q-p.gt.-sqrt(2/9)) qnew = + (11/12)*q - ( 7/12)*p -  7*d
if(q-p.gt.-sqrt(2/9)) pnew = - ( 7/12)*q + (11/12)*p -  1*d
[ Nonequilibrium Baker Map N2 ]
\end{verbatim}
Here and just below the constant {\tt d} is $\sqrt{1/72}$. The inverse mapping N2$^{-1}=$ T{\tt *}N2{\tt *}T
is similar :
\begin{verbatim}
if(q+p.le.-sqrt(2/9)) qnew = + (11/ 6)*q + ( 7/ 6)*p + 14*d
if(q+p.le.-sqrt(2/9)) pnew = + ( 7/ 6)*q + (11/ 6)*p + 10*d
if(q+p.gt.-sqrt(2/9)) qnew = + (11/12)*q + ( 7/12)*p -  7*d
if(q+p.gt.-sqrt(2/9)) pnew = + ( 7/12)*q + (11/12)*p +  1*d
[ Inverse of the Nonequilibrium Map N2 ]
\end{verbatim}
The N2 mapping, shown in {\bf Figure 2}, was selected to model the steady-state fractal structures discovered with
nonequilibrium thermostated molecular dynamics\cite{b3,b4}.  N2 and the slightly more complex N3 embody five
characteristics of those simulations.

The maps are [ 1 ] deterministic, [ 2 ] time-reversible, [ 3 ] chaotic, [ 4 ] fractal, and [ 5 ] dissipative. Here
steady-state dissipation corresponds to a reduction in the variety of available states -- area reduction in the $(q,p)$
state space.  This loss averages a factor of $2^{1/3}$ per iteration ( or ``timestep'' ), with area increasing only
one-third the time and decreasing the remaining two-thirds.  Surprisingly, with the same initial condition,
$(q,p)=(1,0)$, the resulting single-precision periodic orbits generated by N2 and T{\tt *}N2{\tt *}T are considerably
longer, 1,042,249 iterations for both of them, than are those for E and T{\tt *}E{\tt *}T, suggesting ( incorrectly )
that the nonequilibrium situation corresponds to more, rather than fewer, states. The double-precision periodic length
resulting from $(q,p)= (0,0)$ is indeed smaller, at 2,148,754,336,529, than the three-trillion-plus orbit of the
equilibrium map.

Both the Equilibrium and Nonequilibrium Baker maps provide Lyapunov-unstable stretching in the direction perpendicular
to the line $q=p$.  In the parallel direction E, N2, and N3 are all compressive. With repeated iterations they can generate
fractal distributions  which are everywhere discontinuous.  For these maps the compressive motion is modeled perfectly
and exactly by analogous one-dimensional bounded random walks.  Though stochastic, irreversible, and lacking inverses,
the one-dimensional walks are perfect models for the chaos, fractal character, and dissipation seen in our nonequilibrium
maps.  Next we describe these equilibrium and nonequilibrium walks.

\subsection{Parallel Lesson from Stochastic Random Walks}

In addition to its simplicity the chaotic and fractal aspects of the two-dimensional N2 map are shared with an
equivalent one-dimensional stochastic random walk\cite{b5} :
$$
\textstyle{
0 < {\cal R} < \frac{2}{3} \to y = \frac{y}{3} \ ; \ \frac{2}{3} < {\cal R} < 1\to y = \frac{1+2y}{3} \ . \ [ \ {\rm Nonequilibrium \ Walk} \ ] \ .
}
$$
Here ${\cal R}$ is a random number selected from the interval $0 < {\cal R} < 1$. The one-dimensional variable
$y$ is related to dependent variables $(q,p)$ of the nonequilibrium Baker map N2 by :
$$
y=\{ \ [ \ (q+p)/\sqrt{2} \ ] + 1 \ \}/2 \ ;
$$
where the variables are confined to a one-dimensional interval and a two-dimensional diamond-oriented square :
$$
0 \ < \ y \ < \ 1 \ ; \ -\sqrt{2} \ < \ (q,p) \ < \ \sqrt{2} \ .
$$
The ``equilibrium'' version of the random walk E, with left and right steps equally probable, is :
$$
\textstyle{
0 < {\cal R} < \frac{1}{2} \to y = \frac{y}{2} \ ; \ \frac{1}{2} < {\cal R} < 1 \to y = \frac{1+y}{2} \ .
\ [ \ {\rm Equilibrium \ Walk} \ ] \ .
}
$$
The nonequilibrium Walks and Maps generate fractional-dimensional ``fractal'' structures which are ergodic, so
that there is some density arbitrarily close to any point in the walks' unit interval or within the diamond-shaped
2 $\times$ 2 squares of {\bf Figures 1, 2, and 3}.

\subsection{R\'enyi's Information Dimension $D_I$}

R\'enyi's ``information dimension''\cite{b5,b6} $D_I$ quantifies the small-scale structure of fractals like these.
$D_I = D_I(\epsilon = 0)$ is the small-$\epsilon$ limiting ratio of two sums over similar bins spanning the entire
fractal structure.
$$
D_I(\epsilon) \equiv \langle \ \ln(P) \ \rangle/\ln \epsilon \ ;
\ \langle \ \ln(P) \ \rangle \equiv \sum_{\rm bins} P\ln (P)/\sum_{\rm bins} P \ .
$$
The $\{ \ P \ \}$ are the probabilities contained in all of the same-sized bins of size $\epsilon$.

Apart from edge effects, spanning a homogeneous two-dimensional object with area unity using a square grid of
$\epsilon \times \epsilon$ bins generates $\epsilon^{-2}$ such bins with probabilities of $\epsilon^2$ so
that $D_I = 2$. Likewise, spanning the unit interval homogeneously gives $\epsilon^{-1}$ bins of width and
probability $\epsilon$ giving $D_I=1$. The information dimension for the nonequilibrium fractal from N2 shown in
{\bf Figure 3} has been variously estimated\cite{b7} to be 1.7337, $1.741_5$, and 1.7897 implying estimates
for the random-walk information dimension of 0.7337, $0.741_5$, and 0.7897. For more details see our recent arXiv
report\cite{b8}. These differing estimates for $D_I$ are mysteries deserving of cogent explanations.  We are
seeking help to uncover and resolve these contradictory results through the medium of the 2020 Snook Prize Problem.
Let us consider the details of the three estimates for $D_I$ one by one.

\section{Estimating the Random Walk $D_I$ with Cantor-like Mappings}

\begin{figure}
\includegraphics[width=2.0 in,angle=-0.]{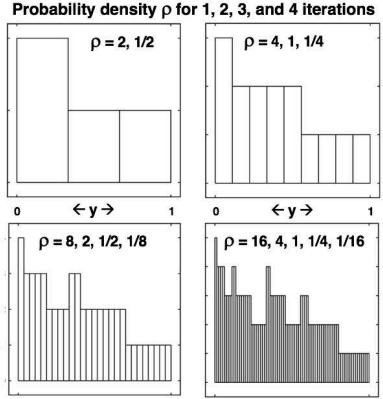}
\caption{
Iterating the Cantor-like Random Walk mapping of a uniform probability eventually leads to a cross section of the
multifractal attractor shown at the left in {\bf Figure 3}.  In both the N2 and N3
nonequilibrium Walks and Baker Maps the probability density is discontinuous everywhere (in the compressive dissipative
$y$ direction parallel to the line $q = p)$ in {\bf Figures 1, 2, and 3}. The distributions shown here, for one,
two, three, and four iterations of the N2 mapping, equivalent to the corresponding finite nonequilibrium N2 random walks,
all have exactly the same information dimensions of 0.78969, disagreeing with the pointwise analyses illustrated in
{\bf Figure 5}.
}
\end{figure}

Doyne Farmer's 1982 review, ``Information Dimension and the Probabilistic Structure of Chaos''\cite{b5} is a short
primer handbook on chaotic dynamics.  He describes the ``fractal dimension'' [ or capacity ] and information
dimension for a wide variety of maps. The fractal dimension is appropriate to fractals with holes, and measures
the dependence of the number of occupied bins on the bin size $\epsilon$. Because our Baker Maps and random walks are
ergodic their ``fractal dimensions'' ( or ``capacities'' ) are 2 and 1 respectively. These capacity dimensions are
integers while the information and correlation dimensions are fractional and so clearly are more useful descriptors.
The correlation dimension describes the dependence of the number of {\it pairs} of points within a bin of size
$\epsilon$ in the limit of vanishing bin size :
$$
D_C = \langle \ \ln \ [ \ {\rm pairs \ with \ }
(r < \epsilon) \ ] \ \rangle /\ln (\epsilon) \longleftarrow ( \ \epsilon \to 0 \ ) \ .
$$
The data plotted on page 286 of Reference 9 indicate that the correlation dimension for map N2 is $1.60\pm 0.01$
corresponding to 0.60 for the corresponding random-walk dimension.  For more information see slides 7 and 8 in
Lecture 10 of the Kharagpur Lectures on our website, {\tt williamhoover.info}.

Farmer analyzes a variant of the Cantor set construction at each iteration of the mapping. Every time the mapping
redistributes the probability of each one-dimensional bin into three similar bins of one third the width, and with
probabilities $[P_0,P_m,P_0]$, reproducing the traditional ``middle-third'' Cantor-set construction for $P_0 = (1/2)$
and $P_m = 0$.  Farmer's more-general model is also equivalent to our nonequilibrium ( compressible ) random walk
with the alternative probability choices, $[(1/6),(2/3),(1/6)]$.  This symmetric arrangement has the same information
dimension as our own walk's $[ \ (2/3),(1/6),(1/6) \ ]$ because the ordering of the bins makes no contribution to the
``information''. See also Figure (10) of Kumi\^c\'ak's work on the random walk version of the N2 nonequilibrium Baker
Map\cite{b10}. We find, using Farmer's equation (13), that the information dimension for our random walk is
$\ln(27/2)/\ln(27) = 0.789 \ 690 \ 082$. 

In our {\bf Figure 4} we show the results of iterating the $[ \ (2/3),(1/6),(1/6) \ ]$ mapping four times from a
distribution which is initially uniform.  With each iteration we can compute $\sum P\ln P$ for the 3, 9, 27, or
81 equal bins that result. It turns out that each time we evaluate the dimensionality we find the  {\it same}
information dimension for the random-walk fractal, 0.7897, agreeing precisely with Farmer's analysis.  The initial
value, after one iteration, is
$$
[ \ (2/3)\ln(2/3) + (1/6)\ln(1/6) + (1/6)\ln(1/6) \ ]/\ln(1/3) = 0.789 \ 690 \ 082 \ .
$$
Further iterations just repeat this value, Farmer's result. The analogous information dimension for the two-dimensional
nonequilibrium Baker Map N2 of {\bf Figures 1 and 2} is $1.789 \ 690 \ 082$. The three-panel nonequilibrium N3 Map
provides exactly this same result.

\section{Estimating the Baker-Map $D_I$ by Single-Point Iteration}

\begin{figure}
\includegraphics[width=2.0 in,angle=-90.]{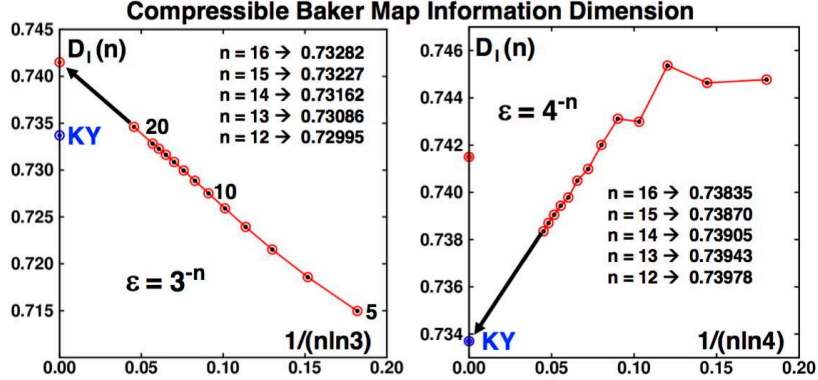}
\caption{
Trillion-point iterations of the nonequilibrium mapping N2, analyzed with a bin size $(1/3)^n$, are consistent
with an equivalent random-walk information dimension $D_I = 0.741_5$. See Reference 7. With a bin size $(1/2)^n$ or $(1/4)^n$
the Kaplan-Yorke conjecture appears more likely to be correct, as indicated by the data at the right. For more details see our
arXiv report 1909.0452.
}
\end{figure}

Rather than mapping {\it distributions} in Cantor-like fashion we can alternatively solve the iterative Baker Map and
Random Walk equations which describe the evolutions of a two-dimensional $(q,p)$ or a one-dimensional $(y)$ {\it point}.
It is practical to analyze a few trillion iterations in a day or so of laptop time.  Spanning the spaces with
grids soon shows that the $(q,p)$ and $(y)$ approaches reliably agree, within the small discrepancies expected
from statistical fluctuations, with the relation.
$$
D_I^{\rm Baker} = D_I^{\rm Walk} + 1 \ .  
$$

The details of the grid-based results soon revealed three flies in the ointment, not just one. The information
dimension requires taking a limit in which the bin size $\epsilon$ vanishes. In practice it is natural to consider the
convergence of series of measurements using bins of size $(1/2)^n$, or $(1/3)^n$, or $(1/5)^n$, ... for
increasing values of $n$. Oddly enough, these series can, and do, disagree.  {\bf Figure 5} compares apparent
information dimensions for two such choices, bin sizes of $(1/3)^n$ and $(1/4)^n$. Because N2 maps two
thirds of the probability into the southwesternmost third of the $(q,p)$ map's domain $(1/3)^n$ is the
``natural'' choice of mesh. Accordingly, our first measurements used $(1/3)^n$ for $n$ as large as 20 and
provided the very nice straight line plot\cite{b7} at the left of the figure.  For the smallest bins we used
{\it trillions} of map iterations to attain four-digit accuracy in $D_I$.  In 1997 we had accepted\cite{b1}
the conventional wisdom\cite{b5,b6} that the Kaplan-Yorke conjecture\cite{b11} was exact for our linear
nonequilibrium map N2. Now, a generation later, we are not so sure !

The results in {\bf Figure 5} show that the two series of bins disagree and that {\it neither} of them agrees with
the Cantor-mapping result, 0.7897. The data are instead consistent with estimates of $0.741_5$ for bin widths $\epsilon$
of $(1/3)^n$ and 0.7337 for $(1/4)^n$.  The latter estimate is the Kaplan-Yorke dimension described in the next
Section. The flies in the ointment are these disparate estimates for what appears to be a well-defined property
of the nonequilibrium fractals, 0.7897, $0.741_5$, and 0.7337 for the one-dimensional walk, corresponding to
1.7897, $1.741_5$, and 1.7337 for the two-dimensional nonequilibrium Baker Map. Maybe mapping distributions and
points are not quite the same ?  Maybe the discontinuous nature of the probability in the $y$ direction is to
blame ? Maybe the two-to-one ratio of the ``natural'' bin sizes of the Cantor-like Mapping is responsible ?

The problem is definitely not a lack of data. This is an advantage of the $(q,p)$ maps over their Cartesian analogs,
which produce relatively short periodic orbits\cite{b7,b8}. Eventually a convergent digital computer's mapping must
repeat.  This is no problem for the one-dimensional $y$ maps as the cycle length for the {\tt Random}\_{\tt Number}
FORTRAN generator we use is said to be greater than $10^{77}$. We verified the lack of any repetitions for $10^{13}$
iterations.  Our double-precision iteration of the $(q,p)$ N2 map, starting at $(0,0)$, eventually settled onto a
periodic orbit of more than two trillion iterations, 2,148,754,336,529 to be precise, and significantly less than
the three trillion plus iterations in the equilibrium periodic orbit in double precision, 3,412,524,575,046. The
dependence of the periodic-orbit length on computational precision can be understood semiquantitatively in terms of
the correlation dimension\cite{b12,b13}, the limiting small-$r$ small-$\epsilon$ power-law describing the number of
pairs of attractor points $\#(r)$ within a distance $\epsilon$, $D_C = \ln[ \ \#(r < \epsilon) \ ]/\ln (\epsilon)$.

\section{Estimating the Information Dimension {\it via} Kaplan-Yorke}

\begin{figure}
\includegraphics[width=2.0 in,angle=-90.]{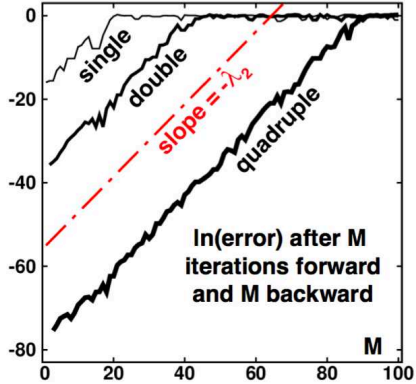}
\caption{
The rms error $\sqrt{q^2+p^2}$ incurred by single-, double-, and quadruple-precision mappings with $M$ steps
forward, using N2, followed by $M$ steps backward, using the inverse mapping  N2$^{-1}$. The overall net error growth
is described well by the red dashed line, with slope $-\lambda_2 = 0.8676$.
}
\end{figure}

Kaplan and Yorke\cite{b11} suggested that the information dimension for typical maps, perhaps including ours,
should be given by their ``conjecture'': $D_I^{\rm KY} = D_\lambda = 1 - (\lambda_1/\lambda_2)$.  Here $\lambda_1$
gives the rate at which two nearby $(q,p)$ points separate under the action of the Baker Map. $\lambda_1 + \lambda_2$
gives the rate at which a small area changes with time, $-(M/3)\ln(2)$, on average, for $M$ iterations, decreasing
two-thirds of the time by a factor two and increasing one-third of the time by the same factor. The two-dimensional
nonequilibrium Baker mapping N2 stretches two-thirds of the measure by a factor of (3/2) perpendicular to the $q=p$
direction while shrinking threefold in the parallel direction.  See again the leftmost panel of {\bf Figure 3}.
The remaining third of the measure stretches threefold perpendicular to $q=p$ and shrinks by a factor (3/2) in the
parallel direction. The resulting Lyapunov exponents for the N2 mapping are :
$$
\lambda_1 = (2/3)\ln(3/2) + (1/3)\ln(3) = (1/3)\ln(27/4) = +0.636 \ 514 \ 168 \ ,
$$
from the stretching, $\propto e^{\lambda_1t}$ and
$$
\lambda_2 = (2/3)\ln(1/3)+ (1/3)\ln(2/3) =(1/3)\ln(2/27) = -0.867 \ 563 \ 228
$$
from the shrinking, $\propto e^{\lambda_2t}$.

To see the effect of the exponential Lyapunov instability of the map we follow a trajectory starting at
$(q,p)=(0,0)$, iterating forward for $M$ iterations, followed by $M$ iterations of the inverse map
N2$^{-1}$ detailed on page 5.

{\bf Figure 6} incorporates the actions of both N2 and N2$^{-1}$.  Starting at the origin and iterating for $M$
steps forward with N2 and then iterating backward, with N2$^{-1}$, likewise for $M$ steps, the growth of the 
roundoff error (due to Lyapunov instabiity) can be measured numerically. The figure shows how this error
varies with $M$. The Lyapunov-unstable perturbation grows exponentially with the elapsed time.  The rate of
growth measured as a function of the  reversal time $M$ shows a slope of $-\lambda_2$. Some thirty years ago
we showed, along with Harald Posch\cite{b14}, that the largest Lyapunov exponent backward in time was simply
the negative of the last going forward in time, in this case $-\lambda_2$, in agreement with the figure. The
simplicity found here, for our reversible maps, does not carry over to flows, where the fluctuations in the
Lyapunov exponents are typically ``large'' relative to $\lambda_1$.

\section{The Three-Panel N3 (x,y) and (q,p) Maps}

Totally by accident, in November 2019 we discovered the N3 map, a somewhat more symmetric and definitely more
conventional three-panel version of the paradoxical N2 mapping.  See again {\bf Figures 2 and 3}. In Cartesian
form the N3 map can be propagated as follows :
\begin{verbatim}                                                                                                             
if (x.lt.-2/3)                  xnew = 6*x + 5                                                                          
if (x.lt.-2/3)                  ynew = (y+2)/3                                                                         
if((x.ge.-2/3).and.(x.le.-1/3)) xnew = 6*x + 3                                                                          
if((x.ge.-2/3).and.(x.le.-1/3)) ynew = (y+0)/3                                                                         
if (x.gt.-1/3)                  xnew = (3*x-1)/2                                                                        
if (x.gt.-1/3)                  ynew = (y-2)/3
[ Cartesian Nonequilibrium N3 Map ]                                                                         
\end{verbatim}

\subsection{Conversion to (q,p) Phase Space}
The Cartesian version of the N3 Map is not only irreversible.  It is also prone to brief fixed cycles.  For instance,
propagating an initial point at the origin shortly produces the cycle
$$                                                                                                                           
{\tt (0,-0.25)} \longleftrightarrow {\tt (-0.50,-0.75)} \ .                                                                  
$$
Such problems are avoided by rotating the $(x,y)$ map 45 degrees (see again {\bf Figure 2}) :
$$                                                                                                                           
{\textstyle                                                                                                                  
q = \sqrt{\frac{1}{2}}(y+x) \ ; \ p = \sqrt{\frac{1}{2}}(y-x) \ \longleftrightarrow \                                        
x = \sqrt{\frac{1}{2}}(q-p) \ ; \ y = \sqrt{\frac{1}{2}}(q+p) \ .                                                            
}                                                                                                                            
$$
Both the instabilities just mentioned are not present in the $(q,p)$ version of the map !

\subsection{Three-Panel N3 (q,p) Map}

Conversion of $(x,y)$ to $(q,p)$ and $(x_{\rm new},y_{\rm new})$ to $(q_{\rm new},p_{\rm new})$ provides 
the more useful and time-reversible N3 mapping suitable for computation where now the constant {\tt d} is
$\sqrt{\frac{1}{72}}$ :
\begin{verbatim}                                                                                                             
if (q-p.lt.-8*d)                    qnew = +19*q/ 6 - 17*p/ 6 + 34*d                                                        
if (q-p.lt.-8*d)                    pnew = -17*q/ 6 + 19*p/ 6 - 26*d                                                        
if((q-p.ge.-8*d).and.(q-p.le.-4*d)) qnew = +19*q/ 6 - 17*p/ 6 + 18*d                                                        
if((q-p.ge.-8*d).and.(q-p.le.-4*d)) pnew = -17*q/ 6 + 19*p/ 6 - 18*d                                                        
if (q-p.gt.-4*d)                    qnew = +11*q/12 -  7*p/12 -  7*d                                                        
if (q-p.gt.-4*d)                    pnew = - 7*q/12 + 11*p/12 -  1*d                                                        
[ Rotated Nonequilibrium N3 Map ]
\end{verbatim}
Though we do not show all the details here the N3 Nonequilibrium mapping behaves very differently to its
close relative N2.  All three versions of the information dimension for N3 appear to give the same result,
$D_I = 1.78969$ for the mapping and 0.78969 for the corresponding random walk :
$$
\textstyle{
          0 < {\cal R} < \frac{1}{6} \to y = \frac{y+2}{3} \ ; \
\frac{1}{6} < {\cal R} < \frac{1}{3} \to y = \frac{y+1}{3} \ ; \
\frac{1}{3} < {\cal R} < 1           \to y = \frac{y+0}{3} \ .
}
$$
\begin{verbatim}
[ Nonequilibrium N3 Walk ]
\end{verbatim}
The fractal generated by this map is particularly simple.  The pattern observed for $0 < y < \frac{1}{3}$
is repeated, but at one fourth the density, for both panels $\frac{1}{3} < y < \frac{2}{3}$ and
$\frac{2}{3} < y < 1$.

\section{Summary and Exhortation}

The existence of three different information dimensions for the simple N2 model ( while the similar N3
model behaves completely conventionally ) is unacceptable from the
pedagogical standpoint. It is conceivable, but hard to contemplate, that many iterations of the maps and
the (equivalent) walks could produce different steady states.  It is likewise hard to believe that the
scaling of an infinite number of iterations could differ from that of a ``large'' number such as $10^{77}$.
Possible but strange! Studies of the maps and their equivalent random walks are bound to uncover the mechanisms
for the three different versions of the information dimension just described. We urge readers to consider
this opportunity to advance our understanding of nonuniform convergence for a ``well-known'' problem area from
more than 80 years ago.  The explanations will surely be both illuminating and stimulating.

\section{Acknowledgment}

We thank several colleagues for help with and criticisms of our investigations: Carl Dettmann, Thomas Gilbert,
Ed Ott, Harald Posch, and James Yorke.  Carl and Thomas were particularly generous with their time and effort
on our behalf.

\end{document}